\documentclass[amsthm,amsmath]{jdsart}
\usepackage{booktabs}
\volume{0}
\issue{0}
\pubyear{2022}
\firstpage{1}
\lastpage{16}
\articletype{research-article}
\aid{JDS1050}
\doi{10.6339/22-JDS1050}

\startlocaldefs

\theoremstyle{plain}

\theoremstyle{remark}
\theoremstyle{definition}
\renewcommand\footnoterule{\kern-3pt \hrule \kern 2.6pt}
\endlocaldefs

\begin{document}
\begin{frontmatter} 
%
%
\pretitle{Data Science in Action}

\title{Privacy-Preserving Inference on the Ratio of Two Gaussians Using Sums}
\runtitle{Privacy-Preserving Inference on the Ratio of Two Gaussians Using Sums}
\begin{aug}
\author[1]{\inits{J.}\fnms{Jingang} \snm{Miao}\thanksref{c1}\ead[label=e1]{jingangmiao@gmail.com}}
\author[1]{\inits{Y.}\fnms{Yiming Paul} \snm{Li}}
\thankstext[type=corresp,id=c1]{Corresponding author.}
\address[1]{\institution{Meta Platforms, Inc.}, \cny{United States}}
\end{aug}
%
%

\begin{abstract}
The ratio of two Gaussians is useful in many contexts of statistical inference.
We discuss statistically valid inference of the ratio under Differential
Privacy (DP). We use the delta method to derive the asymptotic distribution
of the ratio estimator and use the Gaussian mechanism to provide (epsilon,
delta)-DP guarantees. Like many statistics, quantities involved in the
inference of a ratio can be re-written as functions of sums, and sums are
easy to work with for many reasons. In the context of DP, the sensitivity
of a sum is easy to calculate. We focus on getting the correct coverage
probability of 95\% confidence intervals (CIs) of the DP ratio estimator.
Our simulations show that the no-correction method, which ignores the DP
noise, gives CIs that are too narrow to provide proper coverage for small
samples. In our specific simulation scenario, the coverage of 95\% CIs
can be as low as below 10\%. We propose two methods to mitigate the under-coverage
issue, one based on Monte Carlo simulation and the other based on analytical
correction. We show that the CIs of our methods have much better coverage
with reasonable privacy budgets. In addition, our methods can handle weighted
data, when the weights are fixed and bounded.
\end{abstract}


\received{\smonth{4} \sday{14}, \syear{2022}}
\accepted{\smonth{5} \sday{21}, \syear{2022}}
%
%
\end{frontmatter}

\section{Introduction}
\label{sec:intro}

Ratios are used in many types of statistical analyses. Examples include
the ratio of regression coefficients \citep{hirschberg2007providing}, the
therapeutic safety ratio \citep{dunlap1986confidence}, and the percent
difference of the outcome metric between two arms in randomized experiments
\citep{deng2018ApplyingDelta}. The use case we focus on is model calibration.
Being well calibrated is widely regarded as a desirable characteristic
of a classification model \citep{degroot1983comparison}. A model is said
to be calibrated if the predicted scores ($s$) match the average of true
labels ($y$). Specifically, among observations with prediction $s$, the
actual percent of positive labels is equal to $s$ for all values of
$s$. This is intuitive: If a weather forecaster predicts the chance of
rain is 80\%, then we expect to observe rain about 80 out of 100 such predictions
for the forecaster to be considered valid
\citep{miller1962statistical} or reliable \citep{murphy1972scalar}. In
practice, calibration curves are often used to visually check how calibrated
a model is, where the observations are bucketed into $K$ (usually 5, 10,
or 20) groups by model score ($s$), and then the average of $y$ for each
group is plotted against the average of model score $s$. A well calibrated
model will have all points close to the 45-degree line. Equivalently, we
want each ratio of average $s$ and average $y$, which we call the calibration
ratio, to be close to 1.

Statistical analysis has started to face another requirement --- privacy
protection. Online privacy, in particular, has become front and center
for many organizations' analytical tasks \citep{abowd2019census}. Organizations
and corporations are exploring potential solutions to preserve analytical
functionalities while preserving user privacy. Differential Privacy (DP)
has become one of the more popular formal definitions of privacy
\citep{dwork2006calibrating}, which can be achieved by adding random noise.
DP by noise addition comes in two general variants: local DP
\citep{kasiviswanathan2008WhatCan}, where random noise is added to the
individual input data points, and central/global DP, where noise is
added to the intermediate or final output. For example, Google uses local
DP to collect the Chrome web browser's usage data
\citep{erlingsson2014RAPPORRandomized}. Meta (formerly Facebook) has shared
its plan to assess fairness in relation to race in the U.S. in privacy-preserving
ways via a combination of Secure Multiparty Computation (SMPC) and global
DP \citep{meta2021race}.

In a fast-growing literature on differentially private confidence intervals,
there are two main approaches. One approach relies on distributional assumptions
(either directly or via large sample theories), and the other approach
uses resampling and simulation methods to numerically approximate the sampling
distribution of estimators. Here we mention a small subset of works in
the field. \cite{dorazio2015differential} examined the DP confidence interval
for the difference-in-means estimator, and derived the sensitivity for
the standard error of difference-in-means to avoid adding noise separately
to intermediate summary statistics.
\cite{movahedi2021PrivacyPreservingRandomized} describes an industry deployment
in a randomized controlled experiment setting, also focusing on difference-in-means
estimator, but using an alternative approach that is based on noisy intermediate
sufficient statistics and approximate sampling distribution.
\cite{vu2009DifferentialPrivacy},
\cite{gaboardi2016DifferentiallyPrivate}, and
\cite{awan2020DifferentiallyPrivate} study DP hypothesis testing for multinomial
data and binomial data. \cite{karwa2017finite} shows how to construct conservative
DP confidence intervals under normality without knowing the bounds in advance,
but the resulting confidence intervals are usually too wide to be practically
useful. \cite{du2020differentially} and \cite{ferrando2020} improve upon
\cite{karwa2017finite} using simulation to get practical confidence intervals
for the mean estimation problem under normality.
\cite{brawner2018bootstrap} use bootstrapping to compute DP confidence
interval along with a point estimate, without additional privacy budget
under Concentrated DP. \cite{covington2021unbiased},
\cite{evans2019statistically}, and
\cite{evans2021BayesianExperimentation} are some recent efforts to provide
unbiased DP inference to offset the biases by some DP procedures such as
winsorization.

To the best of the authors' knowledge, there is no existing work on Differentially
Private statistical inference on the ratio of two random Gaussian variables.
This work is an attempt to fill this gap. We propose methodology to conduct
statistically valid inference of ratio estimators under DP, with a specific
focus on preventing under-coverage of confidence intervals. We also examine
the case when the data is weighted (e.g., in a complex survey design).
Our methods apply as long as the conditions of using the delta method is
satisfied (see \cite{deng2018ApplyingDelta} for a recent discussion).

\section{Definitions and Methodology}
\label{sec2}

This section defines the quantity of interest and the privacy semantics.
We use $n$ for sample size, $y$ for the label, and $s$ for the score (probability
prediction from a classification model). Both $y$ and $s$ are non-negative.
Further, $l_{y}$, $u_{y}$, $l_{s}$, $u_{s}$ are the lower and upper bounds on
$y$ and $s$ respectively. We focus on the binary classification models,
where the bounds on $y$ and $s$ are $[0, 1]$.\looseness=-1

When the data is weighted, we use $l_{w}$, $u_{w}$ for the lower and upper
bounds of $w$, the sample weights, which are assumed to be fixed (e.g.,
design weights). We also assume that $u_{w}$ is known, which is the case
for example when the bounds are specified in the weight calibration step.

\subsection{Calibration Ratio}
\label{sec2.1}

Given a model, the calibration ratio is simply
$r = \mu _{s} / \mu _{y}$, where $\mu _{s}$ and $\mu _{y}$ are the true
means of $s$ and $y$. An estimator of $r$ is
$\hat{r} = \bar{s} / \bar{y}$. Note this estimator is statistically biased,
but its bias is of order $1/n$ and vanishes quickly as sample size increases.
What's more interesting is its variance.

We will use the fact that $\bar{s} / \bar{y} = \sum s / \sum y$ to use
ratio of sums instead of means, since the ratio of sums is easier to work
with for inference. With a slight abuse of notation, we use
$\bar{s}$ and $\bar{y}$ to denote both the means when data is not weighted
as well as the weighted means when data is weighted.

\subsection{Differential Privacy}

\label{sec2.2}

DP has grown to be one of the most influential privacy definitions in recent
years. In this section we introduce the basic privacy semantics, definitions,
and properties of DP. In this paper we focus on the classical Pure DP definition
($\epsilon $-DP) and Approximate DP definition, also known as ($
\epsilon $, $\delta $)-DP. For a more complete treatment we refer readers
to \cite{dwork2014algorithmic}.

A randomized algorithm satisfies the requirement of DP
\citep{dwork2006calibrating} if for every two neighboring datasets that
differ on exactly one record, and for every possible output, the probabilities
of the output is close up to a multiplicative factor of
$e^\epsilon \approx 1 + \epsilon $ whether the randomized algorithm is
applied on one dataset or the other. This is often called
$\epsilon $-DP or pure DP.

As we can see from the informal definition above, DP requires that the
neighboring datasets result in essentially indistinguishable distributions
of data releases; or more succinctly, close datasets have close outputs.
This requires formal measures for 1) the distance between two datasets,
and 2) the distance between two distributions of output. The choice of
these two distance relations defines the flavor of DP.

There are two popular notions of neighboring datasets in the DP literature.
One is called ``add/remove-one,'' where we can get the neighboring dataset
by either adding or removing one observation. The other one is called ``change-one,''
where we get the neighboring dataset by changing the value of an observation,
instead of adding/removing it to/from the dataset. The change-one definition
can be seen as the result of removing one and then adding another observation
(or in the reverse order). In this paper we use the ``add/remove-one''
definition of neighboring, because we intend to protect the sample sizes
as well, in order to prevent certain privacy attacks such as membership
inference attacks or tracing attacks \citep{Dwork2017}.

\citet{dwork2006our} relaxes the DP requirement by allowing for the violation
of $\epsilon $-DP with a (cryptographically) small probability
$\delta $. This is often called ($\epsilon $, $\delta $)-DP or Approximate
DP. Formally, a randomized algorithm
$M: \mathcal{X}^{n} \rightarrow \mathcal{Y}$ is ($\epsilon $, $\delta $)-DP
if for all neighboring datasets $X, X' \in \mathcal{X}^{n}$ and all outcomes
$T \subseteq \mathcal{Y}$ we have
$\text{Pr}\left (M(X) \in T \right ) \leq e^{\epsilon} \text{Pr}\left (M(X')
\in T \right ) + \delta $.

Two properties of DP algorithms are relevant to this paper
\citep{dwork2014algorithmic}:
\begin{enumerate}
\item \textbf{Closure under composition}: The composition of $K$ differential
private mechanisms, where the $k$th mechanism is
$(\epsilon _{k}, \delta _{k})$-DP, for $1 \leq k \leq K$, is
$(\sum _{k=1}^{K} {\epsilon _{k}}, \sum _{k=1}^{K}{\delta _{k}})$-DP. This
is known as basic composition, which we use in this paper. There are more
advanced theorems that have tighter composition bounds than the basic composition
\citep{kairouz2017CompositionTheorem}.
\item \textbf{Immune to post-processing}: If an algorithm is ($
\epsilon $, $\delta $)-DP, then any post-processing of its outputs (i.e.,
without going back and looking at the raw data again) is still ($
\epsilon $, $\delta $)-DP.
\end{enumerate}

DP provides strong privacy guarantee for the worst-case scenario, at the
cost of utility degradation. The privacy guarantee holds no matter how
the data is distributed and what type of attack happens, but the added
noise makes the statistical inference less precise.

DP makes intuitive sense for robust predictive modeling or statistical
inference \citep{dwork2009DifferentialPrivacy}. The ultimate goal of a
predictive model is to have accurate predictions out of sample, not in
sample. Similarly, the ultimate goal of statistical inference is to generalize
the conclusion beyond the sample at hand. As a result, a small change in
the sample, or one observation in the DP case, should not change the model
or the inference much.

\subsection{Inference}
\label{sec2.3}

For inference, the point estimate of the ratio is simply the ratio of the
two (weighted) means, which is biased but the bias goes away quickly as
sample size increases. So, we instead focus on the confidence interval
(CI), usually at the 95\% confidence level. Due to the Central Limit Theorem,
both the numerator and the denominator of $\hat{r}$ are means of independent
and identically distributed variables and are thus asymptotic Gaussians.
For a ratio of two Gaussians, the delta method shows that the asymptotic
distribution of $\hat{r}$ is itself a Gaussian with variance
%
\begin{eqnarray}
\text{Var}(\hat{r}) = \frac{1}{\mu _{\bar{y}}^{2}} \sigma ^{2}_{
\bar{s}} -2\frac{\mu _{\bar{s}}}{\mu _{\bar{y}}^{3}}\sigma _{\bar{y}
\bar{s}} + \frac{\mu _{\bar{s}}^{2}}{\mu _{\bar{y}}^{4}}\sigma ^{2}_{
\bar{y}},
\label{eq:var00}
\end{eqnarray}
where $\mu _{\bar{s}}$ and $\mu _{\bar{y}}$ are the means of
$\bar{s}$ and $\bar{y}$, $\sigma ^{2}_{\bar{s}}$ and
$\sigma ^{2}_{\bar{y}}$ are their variances, and
$\sigma _{\bar{y}\bar{s}}$ is their covariance. See
\cite{casella2002StatisticalInference} for a derivation.

As a result of the fact that both $y$ and $s$ are non-negative, the distribution
of $r$ is often right skewed. However, the CIs constructed using the delta
method is symmetric. As a result, people sometimes either directly use
$\text{log}{(r)}$ or construct CI for $\text{log}{(r)}$ and exponentiate
both limits back to the original scale. The asymptotic variance of
$\text{log}{(\hat{r})}$ can also be constructed using the delta method:
\begin{eqnarray}
\text{Var}(\text{log}{(\hat{r})}) = \frac{1}{\mu _{\bar{s}}^{2}}
\sigma ^{2}_{\bar{s}} -2\frac{1}{\mu _{\bar{s}} \mu _{\bar{y}}}
\sigma _{\bar{y}\bar{s}} + \frac{1}{\mu _{\bar{y}}^{2}}\sigma ^{2}_{
\bar{y}},
\nonumber
\end{eqnarray}
where the quantities needed are the same as in Equation~\eqref{eq:var00}. In the rest of the paper, we will focus on the ratio
scale and only briefly discuss the log scale in the Simulations and Results
sections.

\subsubsection{DP Mechanism}
\label{sec2.3.1}

In statistics, many quantities of interest can be written as functions
of sums, a fact we make use of here. In particular, sums are appealing
in the context of DP because their sensitivity can be easily calculated.
It is straightforward to re-write the plug-in estimator of Equation~\eqref{eq:var00} in terms of sums, where $x$ is a placeholder for either
$s$ or $y$:
%
\allowdisplaybreaks
\begin{align}
\widehat{\mu _{\bar x}} &=
\frac{\sum _{i=1}^{n} w_{i} x_{i}} {\sum _{i=1}^{n} w_{i}}
\label{eq:mean}
\\
\widehat{\sigma ^{2}_{\bar x}} &=
\frac{\sum _{i=1}^{n} w_{i}^{2}}{(\sum _{i=1}^{n} w_{i})^{2}} \left
\{\frac{\sum _{i=1}^{n}w_{i} x_{i}^{2}}{\sum _{i=1}^{n} w_{i}} -
\left [ \frac{\sum _{i=1}^{n} w_{i} x_{i}}{\sum _{i=1}^{n} w_{i}}
\right ]^{2} \right \}
\label{eq:var_mean}
\\
\widehat{\sigma _{{\bar y} {\bar s}}} &=
\frac{\sum _{i=1}^{n} w_{i}^{2}}{(\sum _{i=1}^{n} w_{i})^{2}} \left
\{\frac{\sum _{i=1}^{n} w_{i} y_{i} s_{i}}{\sum _{i=1}^{n} w_{i}} -
\frac{\sum _{i=1}^{n} w_{i} y_{i} \sum _{i=1}^{n} w_{i} s_{i}}{(\sum _{i=1} w_{i})^{2}}
\right \}
\label{eq:cov_mean}
\end{align}

To be explicit, up to 7 sums are needed: $\sum _{i=1}^{n} w_{i}$,
$\sum _{i=1}^{n} w_{i} y_{i}$, $\sum _{i=1}^{n} w_{i} s_{i}$,
$\sum _{i=1}^{n} w_{i}^{2}$, $\sum _{i=1}^{n} w_{i} y_{i}^{2}$,
$\sum _{i=1}^{n} w_{i} s_{i}^{2}$, and
$\sum _{i=1}^{n} w_{i} y_{i} s_{i}$. However, for a binary classification
model, $y$ is either 0 or 1, so
$\sum _{i=1}^{n} w_{i} y_{i} = \sum _{i=1}^{n} w_{i} y_{i}^{2}$, leading
to 6 sums needed. Further, when the data is not weighted, i.e.,
$w_{i}=1$ for all $i$, then
$\sum _{i=1}^{n} w_{i} = \sum _{i=1}^{n} w_{i}^{2}$, leading to only 5
sums needed. Also, you may recognize the inverse of Kish's effective sample
size $(\sum _{i=1}^{n} w_{i})^{2} / (\sum _{i=1}^{n} w_{i}^{2})$
\citep{kish1965survey} in Equations~\eqref{eq:var_mean} and~\eqref{eq:cov_mean}. Without weights, the effective sample size is simply
$n$. The effective sample size indicates the loss of efficiency due to
weighting.

Recall that one reason we use the sums is that their sensitivity can be
easily obtained. Under the add/remove-one definition of neighboring datasets,
the sensitivity of each sum is simply the summand with $s$, $y$, and
$w$ replaced by their (positive) upper bounds. In the binary classification
case, the bounds for $s$ and $y$ are [0, 1], so the sensitivity for all
sums is simply $u_{w}$.

We use the Gaussian mechanism to achieve ($\epsilon $, $\delta $)-DP
\citep{dwork2006our}, which uses Gaussian noise with standard deviation:
%
\begin{eqnarray}
\sigma = \frac{\Delta \sqrt{2 \text{log}{(1.25/\delta )}}}{\epsilon}
\label{eq:noise_sd}
.
\end{eqnarray}

Improved methods are available so that less noise is needed
\citep{balle2018improving}, where the variance of the noise may have to
be obtained numerically. Here, $\sum _{i=1}^{n} w_{i} y_{i}$, for example,
will be released as
$(\sum _{i=1}^{n} w_{i} y_{i})_{\text{dp}} = \sum _{i=1}^{n} w_{i} y_{i}
+ e$, where we use a subscript dp to indicate the noisy quantity that can
be released. Here, $e$ is the noise term coming from a Gaussian distribution
$e \sim \text{Gaussian}(0, \sigma ^{2}_{\sum _{i=1}^{n} w_{i} y_{i}})$,
where $\sigma $ is obtained by plugging $\Delta = u_{w}$ into Equation~\eqref{eq:noise_sd}. Due to composition, the global budget is split among
quantities released. For example, if 6 sums are released, then each one
would get to use 1/6 of the total privacy budget: ($\epsilon / 6$,
$\delta / 6$). Tighter composition theorems can be used for large number
of composition rounds, but here we use the basic composition for easier
exposition.

In model calibration exercises, multiple testings are common; e.g., across
many models, many subgroups, and many time periods. We focus on the Gaussian
mechanism due to its better utility under a large number of compositions.
For smaller compositions, the Laplace mechanism tends to have better utility.
We include simulations based on the Laplace mechanism in the Appendix and
briefly discuss the conditions under which the Gaussian or Laplace mechanism
is more appropriate. When Laplace noises are added, the numerator and denominator
of the ratio are not longer Gaussians, which violates the assumptions of
the Analytical correction method to be introduced in Section~\ref{ci_analytical}. However, the method seems robust against this violation.

\subsection{CI Calculation}
\label{sec2.4}

Once the DP version of the up to 7 sums are released, all calculations based
on them are post-processing, so the privacy guarantee remains the same,
by the post-processing property of ($\epsilon $, $\delta $)-DP. The point
estimate is simply
\begin{eqnarray*}
\hat{r}=
\frac{\left (\sum _{i=1}^{n} w_{i} s_{i} \right )_{\text{dp}}} {\left (\sum _{i=1}^{n} w_{i} y_{i}\right )_{\text{dp}}}.
\end{eqnarray*}
What's more interesting is its variance. Instead of ignoring the DP noises
added, we propose two methods that appropriately account for them in the
CI calculation.

Once the point estimates and variances are obtained via any of the three
methods below, hypothesis testing of the equality of the two ratios
$r_{1}$ and $r_{2}$ can be easily carried out since
$\hat{r}_{1} - \hat{r}_{2} \xrightarrow{d} \text{Normal}(r_{1} - r_{2},
\sigma ^{2}_{r_{1}} + \sigma ^{2}_{r_{2}})$, where $\hat{r}_{1}$ and
$\hat{r}_{2}$ are the point estimates and $\sigma ^{2}_{r_{1}}$ and
$\sigma ^{2}_{r_{2}}$ are their variances.

\subsubsection{No Correction}
\label{sec2.4.1}

What is often done in practice is simply ignore the DP noise added and
apply no correction. To be explicit, the DP version of the sums are plugged
into Equations~\eqref{eq:mean} and~\eqref{eq:cov_mean} to get the mean
and variance/covariance estimates, which are then plugged into Equation~\eqref{eq:var00} to get the final variance estimate. We call the variance
obtained this way $\sigma ^{2}_{\text{no\_correction}}$, which ignores uncertainty
due to DP noises and thus gives CIs that are expected to be too narrow
in small sample settings.

\subsubsection{Monte Carlo}
\label{sec2.4.2}

To estimate the variance injected by the DP mechanism to the ratio estimate,
we can use Monte Carlo simulations. Recall that the ratio of means is the
same as the ratio of sums. The procedure is as follows:
\begin{enumerate}
\item calculate point estimate
\begin{eqnarray*}
\hat{r} =
\frac{\left (\sum _{i=1}^{n} w_{i} s_{i} \right )_{\text{dp}}} {\left (\sum _{i=1}^{n} w_{i} y_{i}\right )_{\text{dp}}}
\end{eqnarray*}
\item for $b = 1, \ldots, B$, where B is a large integer (e.g., 200):
\begin{enumerate}
\item generate independent Gaussian noises $e_{s, b}$ and $e_{y, b}$ for
$\sum _{i=1}^{n} w_{i} s_{i}$ and $\sum _{i=1}^{n} w_{i} y_{i}$, respectively.
Noises are from distributions with the same variances as in the original
DP mechanism, according to Equation~\eqref{eq:noise_sd}.
\item calculate
\begin{eqnarray*}
\hat{r}_{b}=
\frac{
\left (\sum _{i=1}^{n} w_{i} s_{i} \right )_{\text{dp}} + e_{s, b}
}
{
\left (\sum _{i=1}^{n} w_{i} y_{i}\right )_{\text{dp}} + e_{y, b}
}
\end{eqnarray*}
\end{enumerate}
\item the extra variance due to DP is then estimated as
\begin{eqnarray*}
\sigma ^{2}_{\text{extra}} = \frac{1}{B} \sum _{b=1}^{B} (\hat{r}_{b} -
\hat{r}) ^ {2}
\end{eqnarray*}
\item the final variance estimate is
$\sigma ^{2}_{\text{sim}} = \sigma ^{2}_{\text{no\_correction}} +
\sigma ^{2}_{\text{extra}}$
\end{enumerate}

Note that we are not looking at the raw data beyond the released DP sums
and thus not consuming additional privacy budget due to the post-processing
property of DP. The Monte Carlo method is easy to implement. In addition,
the computation is fairly cheap since it can be vectorized.

The $\sigma ^{2}_{\text{extra}}$ term is not an unbiased estimator: In step
2(b) noises are added to the DP sums, whereas the random noises are added
to the non-DP sums in the DP mechanism that produces the point estimate.
The potential bias decreases with an increasing sample size as the DP sums
approach non-DP sums. In the simulations, we will test the method's robustness
by including cases where the privacy budget is small so that the noise
tends to be big.

\subsubsection{Analytical Correction}
\label{ci_analytical}

Recall from Equation~\eqref{eq:var00} that the variance of $\hat{r}$ depends
on the means and variance/covariance of $\bar{s}$ and $\bar{y}$. For convenience
we again use the ratio of sums instead of means.

How do the Gaussian noises added to $\sum _{i=1}^{n} w_{i} s_{i}$ and
$\sum _{i=1}^{n} w_{i} y_{i}$ change their variance? The noise term is
independent of the true quantity, so the variance of the released quantity,
which is a sum of the two, is simply the sum of the variances. Further,
the independent noises do not change the covariance term. As a result,
all we need is to add the variance of the noise to the variance terms.

We follow the steps below to analytically adjust the variance of the ratio
estimator in Equation~\eqref{eq:var00}:
\begin{enumerate}
\item Plug $\left (\sum _{i=1}^{n} w_{i}\right )_{\text{dp}}$,
$\left (\sum _{i=1}^{n} w_{i} y_{i}\right )_{\text{dp}}$,
$\left (\sum _{i=1}^{n} w_{i} s_{i}\right )_{\text{dp}}$,
$\left (\sum _{i=1}^{n} w_{i}^{2}\right )_{\text{dp}}$,
$\left (\sum _{i=1}^{n} w_{i} y_{i}^{2}\right )_{\text{dp}}$,
$\left (\sum _{i=1}^{n} w_{i} s_{i}^{2}\right )_{\text{dp}}$, and
$\left (\sum _{i=1}^{n} w_{i} y_{i} s_{i}\right )_{\text{dp}}$ into Equations~\eqref{eq:mean} through~\eqref{eq:cov_mean}, to get estimates
$\widehat{\mu _{\bar s}}$, $\widehat{\mu _{\bar y}}$,
$\widehat{\sigma ^{2}_{\bar s}}$, $\widehat{\sigma ^{2}_{\bar y}}$, and
$\widehat{\sigma _{\bar y \bar s}}$.
\item Translate those to the corresponding estimates for sums:
$\widehat{\mu _{\bar s}} \cdot \left (\sum _{i=1}^{n} w_{i}\right )_{
\text{dp}}$,
$\widehat{\mu _{\bar y}} \cdot \left (\sum _{i=1}^{n} w_{i}\right )_{
\text{dp}}$,
$\widehat{\sigma ^{2}_{\bar s}} \cdot \left (\sum _{i=1}^{n} w_{i}
\right )_{\text{dp}}^{2}$,
$\widehat{\sigma ^{2}_{\bar y}} \cdot \left (\sum _{i=1}^{n} w_{i}
\right )_{\text{dp}}^{2}$, and
$\widehat{\sigma _{\bar y \bar s}} \cdot \left (\sum _{i=1}^{n} w_{i}
\right )_{\text{dp}}^{2}$.
\item Analytically correct the variance terms as follows:
\begin{enumerate}
\item
$\widehat{\sigma ^{2}_{\bar s}} \cdot \left (\sum _{i=1}^{n} w_{i}
\right )_{\text{dp}}^{2} + \sigma ^{2}_{\sum _{i=1}^{n} w_{i} s_{i}}$
\item
$\widehat{\sigma ^{2}_{\bar y}} \cdot \left (\sum _{i=1}^{n} w_{i}
\right )_{\text{dp}}^{2} + \sigma ^{2}_{\sum _{i=1}^{n} w_{i} y_{i}}$,
\end{enumerate}
where the added term to each is the variance of the DP noises based on
Equation~\eqref{eq:noise_sd}
\item Plug those corrected terms for the sums in place of the terms for
the means into Equation~\eqref{eq:var00} to get corrected variance estimate.
\end{enumerate}

\section{Simulations}
\label{sec3}

\nocite{harris2020array} With a sample size of $5{,}000$ or
$10{,}000$, we simulated $s \sim \text{Beta}(2, 2)$,
$y \sim \text{Bernoulli}(s/1.1)$ (so that true calibration ratio was
$1.1$), and $w$ as Exponential(1) clipped to the range of [$1/3$,
$3$]. Values of $\epsilon $ used included $\{0.2, 0.5, 1.0, 4.0\}$,
$\delta =$ 1e-6, and both weighted and unweighted data were analyzed. For
many use cases, a calibration ratio of 1.0 corresponds to the null hypothesis.
Here a calibration ratio of 1.1 was used to represent the situation where
the alternative hypothesis is true. We did, however, carry out simulations
with a true calibration ratio of 1.0, based on which the main conclusions
would not change and the width of CIs were narrower than for a value of
1.1.

For each simulated dataset, we generated the 95\% Wald confidence intervals,
obtained the width of the intervals, checked whether each covered the true
(log) calibration ratio, and calculated the interval score (the smaller
the better) using Equation 43 of \citet{gneiting2007strictly} for the following
methods
\begin{itemize}
\item Public: the public method without DP
\item No\_correction: the method without correction for DP noise
\item Monte Carlo: the correction based on Monte Carlo simulation
\item Analytical correction: the correction based on modified variance
terms
\end{itemize}

We also calculated the effective sample size, which gave us a rough idea
of how variable the weights are, using the Kish formula
$(\sum _{i=1}^{n} w_{i})^{2} / (\sum _{i=1}^{n} w_{i}^{2})$
\citep{kish1965survey}. Recall that the inverse of Kish's effective sample
size appeared in Equations~\eqref{eq:var_mean} and~\eqref{eq:cov_mean}. We repeated the simulation $1{,}000$ times. The python
code for the simulation can be found at
\url{https://github.com/miaojingang/private_ratio}.

\begin{table}[t!]
\caption{Average width, coverage, and CI score of 95\protect \%
confidence intervals for the ratio, using the Gaussian mechanism. Public:
no noise added; the results do not change as a function of
$\epsilon $. No correction: ignoring the fact that DP noise was added.
Monte Carlo: correction via Monte Carlo simulation. Analytical: correction
via modified variance terms.}
\label{ratio_tab}
\renewcommand{\arraystretch}{1.05}
\begin{center}
\tabcolsep=0pt
\begin{tabular*}{\textwidth}{@{\extracolsep{\fill}}l ccc ccc ccc}
\toprule
    & \multicolumn{3}{c}{No Correction}
    & \multicolumn{3}{c}{Monte Carlo}
    & \multicolumn{3}{c}{Analytical} \\

   $\epsilon $  & width & coverage & Score
               & width & coverage & score
               & width & coverage & score \\
\midrule
\multicolumn{10}{c}{No weights, $n=5{,}000$, effective $n=5{,}000$} \\
\multicolumn{10}{c} {public method: width $=$ 0.061, coverage $=$ 0.951, score $=$ 0.073} \\
0.2 &      0.060 &        0.231 &              2.074 &     0.370 &       0.945 &             0.452 &     0.367 &       0.943 &             0.451 \\
0.5 &      0.061 &        0.538 &              0.489 &     0.156 &       0.952 &             0.191 &     0.156 &       0.946 &             0.192 \\
1.0 &      0.061 &        0.782 &              0.158 &     0.094 &       0.948 &             0.115 &     0.094 &       0.950 &             0.116 \\
4.0 &      0.061 &        0.935 &              0.077 &     0.064 &       0.943 &             0.076 &     0.064 &       0.942 &             0.076 \\[2ex]

\multicolumn{10}{c}{With weights, $n=5{,}000$, effective $n=3{,}032$} \\
\multicolumn{10}{c} {public method: width $=$ 0.078, coverage $=$ 0.949, score $=$ 0.093} \\
0.2 &      0.090 &        0.076 &             10.296 &     8.783 &       0.949 &             9.040 &     1.532 &       0.939 &             1.852 \\
0.5 &      0.076 &        0.205 &              3.190 &     0.549 &       0.946 &             0.659 &     0.535 &       0.940 &             0.652 \\
1.0 &      0.077 &        0.398 &              1.127 &     0.274 &       0.941 &             0.332 &     0.272 &       0.940 &             0.333 \\
4.0 &      0.078 &        0.867 &              0.139 &     0.101 &       0.951 &             0.121 &     0.101 &       0.949 &             0.121 \\[2ex]

\multicolumn{10}{c}{No weights, $n=10{,}000$, effective $n=10{,}000$} \\
\multicolumn{10}{c} {public method: width $=$ 0.043, coverage $=$ 0.949, score $=$ 0.050} \\
0.2 &      0.043 &        0.354 &              0.826 &     0.185 &       0.956 &             0.217 &     0.185 &       0.954 &             0.215 \\
0.5 &      0.043 &        0.699 &              0.178 &     0.084 &       0.952 &             0.096 &     0.084 &       0.946 &             0.095 \\
1.0 &      0.043 &        0.870 &              0.071 &     0.056 &       0.954 &             0.063 &     0.056 &       0.955 &             0.063 \\
4.0 &      0.043 &        0.945 &              0.051 &     0.044 &       0.951 &             0.051 &     0.044 &       0.951 &             0.051 \\[2ex]

\multicolumn{10}{c}{With weights, $n=10{,}000$, effective $n=6{,}161$} \\
\multicolumn{10}{c} {public method: width $=$ 0.055, coverage $=$ 0.953, score $=$ 0.063} \\
0.2 &      0.054 &        0.126 &              4.351 &     0.699 &       0.956 &             0.806 &     0.669 &       0.952 &             0.784 \\
0.5 &      0.054 &        0.322 &              1.290 &     0.268 &       0.953 &             0.313 &     0.266 &       0.958 &             0.309 \\
1.0 &      0.055 &        0.555 &              0.418 &     0.141 &       0.954 &             0.165 &     0.141 &       0.951 &             0.163 \\
4.0 &      0.055 &        0.910 &              0.075 &     0.064 &       0.952 &             0.072 &     0.064 &       0.952 &             0.072 \\
\bottomrule
\end{tabular*}
\end{center}
\end{table}

The results for ratio estimation are summarized in Table~\ref{ratio_tab}. The public version, as expected, has coverages fairly
close to the nominal level of 95\%. As expected, its CI score is the best
among all methods.

The no-correction method under covers in most cases, and its CIs are similar
to or only slightly wider than those of the public method. This is because
the no-correction method does not account for the extra variability introduced
by the DP mechanism. As a result, its CIs are too narrow, especially for
cases with small sample sizes and/or small privacy budget and/or weighted
sample. For example, on the weighted data with $n=5{,}000$,
$\epsilon =0.2$, its CIs only covers the true value 7.6\% of the time,
which is grossly lower than the nominal coverage level. Its CI scores are
the worst among all methods.

\begin{table}[t!]
\caption{Average width, coverage, and CI score of 95\protect \%
confidence intervals for the log ratio, using the Gaussian mechanism. Public:
no noise added and thus Non-DP; the results do not change as a function
of $\epsilon $. No correction: ignoring the fact that DP noise was added.
Monte Carlo: correction via Monte Carlo simulation. Analytical: correction
via modified variance terms.}
\label{log_ratio_tab}
\renewcommand{\arraystretch}{1.05}
\begin{center}
\tabcolsep=0pt
\begin{tabular*}{\textwidth}{@{\extracolsep{\fill}}l ccc ccc ccc}
\toprule
    & \multicolumn{3}{c}{No Correction}
    & \multicolumn{3}{c}{Monte Carlo}
    & \multicolumn{3}{c}{Analytical} \\

   $\epsilon $  & width & coverage & Score
               & width & coverage & score
               & width & coverage & score \\
\midrule
\multicolumn{10}{c}{No weights, $n=5{,}000$, effective $n=5{,}000$} \\
\multicolumn{10}{c} {public method: width $=$ 0.055, coverage $=$ 0.953, score $=$ 0.066} \\
0.2 &      0.055 &        0.232 &              1.882 &     0.333 &       0.944 &             0.405 &     0.332 &       0.941 &             0.406 \\
0.5 &      0.055 &        0.535 &              0.445 &     0.142 &       0.950 &             0.173 &     0.142 &       0.950 &             0.174 \\
1.0 &      0.055 &        0.783 &              0.143 &     0.086 &       0.950 &             0.105 &     0.086 &       0.952 &             0.105 \\
4.0 &      0.055 &        0.937 &              0.070 &     0.058 &       0.944 &             0.069 &     0.058 &       0.944 &             0.069 \\[2ex]

\multicolumn{10}{c}{With weights, $n=5{,}000$, effective $n=3{,}082$} \\
\multicolumn{10}{c} {public method: width $=$ 0.071, coverage $=$ 0.948, score $=$ 0.085} \\
0.2 &      0.077 &        0.075 &              8.864 &     1.340 &       0.915 &             1.367 &     1.268 &       0.975 &             1.349 \\
0.5 &      0.069 &        0.206 &              2.884 &     0.486 &       0.943 &             0.579 &     0.482 &       0.942 &             0.582 \\
1.0 &      0.070 &        0.395 &              1.023 &     0.247 &       0.943 &             0.300 &     0.247 &       0.943 &             0.301 \\
4.0 &      0.071 &        0.864 &              0.126 &     0.092 &       0.952 &             0.110 &     0.092 &       0.951 &             0.111 \\[2ex]

\multicolumn{10}{c}{No weights, $n=10{,}000$, effective $n=10{,}000$} \\
\multicolumn{10}{c} {public method: width $=$ 0.039, coverage $=$ 0.948, score $=$ 0.046} \\
0.2 &      0.039 &        0.355 &              0.752 &     0.168 &       0.951 &             0.196 &     0.168 &       0.956 &             0.194 \\
0.5 &      0.039 &        0.701 &              0.161 &     0.076 &       0.947 &             0.087 &     0.076 &       0.949 &             0.086 \\
1.0 &      0.039 &        0.873 &              0.064 &     0.051 &       0.955 &             0.058 &     0.051 &       0.957 &             0.057 \\
4.0 &      0.039 &        0.945 &              0.046 &     0.040 &       0.949 &             0.046 &     0.040 &       0.950 &             0.046 \\[2ex]

\multicolumn{10}{c}{With weights, $n=10{,}000$, effective $n=6{,}161$} \\
\multicolumn{10}{c} {public method: width $=$ 0.050, coverage $=$ 0.951, score $=$ 0.057} \\
0.2 &      0.049 &        0.123 &              3.953 &     0.613 &       0.960 &             0.694 &     0.604 &       0.964 &             0.690 \\
0.5 &      0.050 &        0.321 &              1.174 &     0.243 &       0.955 &             0.283 &     0.242 &       0.960 &             0.281 \\
1.0 &      0.050 &        0.556 &              0.380 &     0.128 &       0.952 &             0.150 &     0.128 &       0.955 &             0.148 \\
4.0 &      0.050 &        0.910 &              0.068 &     0.058 &       0.953 &             0.065 &     0.058 &       0.953 &             0.065 \\

\bottomrule
\end{tabular*}
\end{center}
\end{table}

Both correction methods have much better coverage. As $\epsilon $ gets
smaller, more noise is injected by the DP mechanism, and both correction
methods correctly account for that by giving wider CIs that have the right
coverage. The correction methods' CI scores are worse than the public method
but better than the no-correction method. With a large sample size and
a larger privacy budget, the DP CIs are only slightly wider than the public
ones; for example, with $n=10{,}000$, $\epsilon =4.0$ and no weights, both
correction methods have a mean CI width of $0.044$, which is barely larger
than the public method's $0.043$. The CI scores also are virtually the
same as that of the public method. Privacy was preserved almost for free.
On the other hand, the increase in CI width is more pronounced for smaller
sample sizes, smaller privacy budgets, and weighted data. Further, when
the privacy budget is too small relative to the sample size, the methods
could still under cover and the CIs can be too wide. For example, with
$n=5{,}000$, $\epsilon =0.2$ and weighted data, the Monte Carlo method's
coverage is only 91.5\% for the log ratio (Table~\ref{log_ratio_tab}),
and its CIs are too wide to be useful for the ratio (Table~\ref{ratio_tab}). In situations like this, practitioners could explore
larger samples and/or larger privacy budget, in addition to potential optimizations
we enumerate in Section~\ref{discussion}.

For the estimation of the log ratio (Table~\ref{log_ratio_tab}), the comparisons
among the methods are similar to those for the ratio.

\section{Discussion}
\label{discussion}

We explored the ratio estimation problem and proposed a DP mechanism based
on adding noise to summary statistics. We also proposed two variance correction
methods that give statistically valid CIs under DP. Our simulations confirmed that the DP noise should
not be ignored in ratio inference unless the sample size is large and/or
the privacy budget is generous; otherwise, the CIs can be too narrow to
cover the true values at the nominal level. The proposal has a few nice
features. It is simple: The sums are easy to compute, their sensitivity
is trivial to calculate, and the variance corrections to get valid CIs
are straightforward. It is flexible: Suppose the data has a hierarchical
structure. For example, if the inference is done at the state level and
later on one wants to aggregate to national level. The sums can be trivially
added up. It is extensible: The variance correction methods can be extended
to inference on other quantities. Sums are the building blocks of many
statistics, including the moments and in turn some more complex quantities
that depend on the moments. Therefore, DP mechanisms based on noising sums
can be applied to other statistics.

This work represents an early effort on ratio estimation under DP. Further
optimizations may be able to achieve better privacy-utility trade-off.
\cite{balle2018improving} propose an Analytic Gaussian Mechanism that reduces
the noise variance compared to the classical Gaussian mechanism in Equation~\eqref{eq:noise_sd}, especially in the high-privacy ($\epsilon \to 0$)
and high-dimensional regime. Similarly, alternative DP mechanisms such
as truncated Laplace for ($\epsilon $, $\delta $)-DP
\citep{geng2020TightAnalysis} could achieve more precise measurements than
Gaussian mechanisms. In cases when many summary statistics need to be privatized,
advanced composition of privacy loss
\citep{kairouz2017CompositionTheorem} or alternative privacy definitions
such as Renyi DP \citep{mironov2017RenyiDifferential}, zero-concentrated
DP \citep{bun2016}, and Gaussian DP \citep{dong2019Gaussian} can provide
tighter accumulation of privacy loss. In addition, there may be smarter
ways of allocating the privacy budget than evenly splitting the budget
among summary statistics, to improve the utility without incurring additional
privacy cost. In use cases with tight privacy budget and high accuracy
requirements, it may help to release fewer intermediary quantities when
possible so that the each quantity gets a bigger privacy budget.

Simulation and resampling have also been used to account for DP noises.
\cite{du2020differentially, ferrando2020} use simulations to directly measure
the combined uncertainty from sampling and DP noise, as opposed to our
methods that account for DP uncertainty separately from the sampling uncertainty.
Resampling methods, such as non-parametric bootstrapping, have also been
proposed to get the standard error of DP statistics without additional
privacy loss \citep{brawner2018bootstrap}. When sample sizes are huge,
subsampling could also help reduce the computational cost
\citep{kleiner2014ScalableBootstrap}.

Finally, we briefly discuss sampling and weighting. Further privacy amplification
by subsampling is possible in certain use cases. When the dataset is a
sample from a larger dataset, and the individual identities in the sample
are kept secret, we could improve the privacy analysis by subsampling,
known as the privacy amplification by subsampling
\citep{balle2020PrivacyProfiles}. It would be interesting to explore how
sampling weights and different sampling schemes affect privacy in inference.
Another direction is to explore how DP may work with more generic types
of weights that are not necessarily fixed or that are with no known bounds.
One popular example is calibration weights
\citep{deville1992calibration}, which are random since they depend on the
sample at hand.

\begin{appendix}
\setcounter{secnumdepth}{0}
\section{Appendix}
\label{appendix:laplace}

The Laplace mechanism draws random noise from the Laplace distribution
to achieve $\epsilon$-DP guarantee. The probability density function of
the Laplace distribution (centered at 0) with scale $b$ is:
%
\begin{eqnarray}
\text{Lap}\left (x\mid b\right )=\frac{1}{2b}\exp \left (-
\frac{|x|}{b}\right ).
\end{eqnarray}

Given a $L_{1}$ global sensitivity of $\Delta$ and a privacy loss parameter
of $\epsilon$, the Laplace noise is drawn from a Laplace distribution
with scale $\Delta / \epsilon$.

Gaussian noise has a few advantages over Laplace noise: 1) the Gaussian
mechanism calibrates the noise proportional to the $L_{2}$ sensitivity,
which is often much smaller than $L_{1}$ sensitivity used by the Laplace
mechanism in vector-output functions. 2) For the same variance, the Gaussian
distribution's tails decay much faster than the Laplace distribution. 3)
In many applications, other sources of noise or measurement errors are
often (approximately) Gaussian, so Gaussian noise works better due to closure
under addition. 4) Moreover, the Gaussian mechanism tends to work better
under a large number of compositions due to tighter composition theorems.

However, for a small number of queries/compositions, the Laplace mechanism
may have an edge in accuracy: for the same value of $\epsilon$ and typical
values of $\delta$ (which the Laplace mechanism does not depend on), Laplace
noise has smaller variance than Gaussian noise. Another advantage of the
Laplace mechanism is that it achieves $\epsilon$-DP instead of
$\left (\epsilon , \delta \right )$-DP, which may be preferred in some
applications.

\begin{table}[htbp]
\caption{Average width, coverage, and CI score of 95\protect \%
confidence intervals for the ratio, using the Laplace mechanism. Public:
no noise added and thus Non-DP; the results do not change as a function
of $\epsilon $. No correction: ignoring the fact that DP noise was added.
Monte Carlo: correction via Monte Carlo simulation. Analytical: correction
via modified variance terms.}
\label{ratio_tab_laplace}
\renewcommand{\arraystretch}{1.075}
\begin{center}
\tabcolsep=0pt
\begin{tabular*}{\textwidth}{@{\extracolsep{\fill}}l ccc ccc ccc}
\toprule
    & \multicolumn{3}{c}{No Correction}
    & \multicolumn{3}{c}{Monte Carlo}
    & \multicolumn{3}{c}{Analytical} \\

   $\epsilon $  & width & coverage & Score
               & width & coverage & score
               & width & coverage & score \\
\midrule
\multicolumn{10}{c}{No weights, $n=5{,}000$, effective $n=5{,}000$} \\
\multicolumn{10}{c} {public method: width $=$ 0.061, coverage $=$ 0.951, score $=$ 0.073} \\
0.2 &      0.061 &        0.730 &              0.238 &     0.109 &       0.937 &             0.149 &     0.109 &       0.940 &             0.147 \\
0.5 &      0.061 &        0.896 &              0.094 &     0.071 &       0.948 &             0.090 &     0.071 &       0.946 &             0.089 \\
1.0 &      0.061 &        0.936 &              0.078 &     0.064 &       0.947 &             0.077 &     0.064 &       0.947 &             0.077 \\
4.0 &      0.061 &        0.949 &              0.073 &     0.061 &       0.950 &             0.073 &     0.061 &       0.950 &             0.073 \\[2ex]

\multicolumn{10}{c}{With weights, $n=5{,}000$, effective $n=3{,}082$} \\
\multicolumn{10}{c} {public method: width $=$ 0.078, coverage $=$ 0.949, score $=$ 0.093} \\
0.2 &      0.077 &        0.416 &              1.505 &     0.344 &       0.936 &             0.473 &     0.339 &       0.938 &             0.468 \\
0.5 &      0.078 &        0.699 &              0.362 &     0.152 &       0.942 &             0.203 &     0.152 &       0.941 &             0.202 \\
1.0 &      0.078 &        0.853 &              0.150 &     0.102 &       0.938 &             0.127 &     0.102 &       0.939 &             0.127 \\
4.0 &      0.078 &        0.944 &              0.096 &     0.080 &       0.948 &             0.096 &     0.080 &       0.949 &             0.096 \\[2ex]

\multicolumn{10}{c}{No weights, $n=10{,}000$, effective $n=10{,}000$} \\
\multicolumn{10}{c} {public method: width $=$ 0.043, coverage $=$ 0.949, score $=$ 0.050} \\
0.2 &      0.043 &        0.829 &              0.098 &     0.063 &       0.951 &             0.077 &     0.063 &       0.947 &             0.078 \\
0.5 &      0.043 &        0.934 &              0.056 &     0.047 &       0.955 &             0.055 &     0.047 &       0.955 &             0.055 \\
1.0 &      0.043 &        0.946 &              0.052 &     0.044 &       0.953 &             0.052 &     0.044 &       0.952 &             0.052 \\
4.0 &      0.043 &        0.949 &              0.050 &     0.043 &       0.950 &             0.050 &     0.043 &       0.950 &             0.050 \\[2ex]

\multicolumn{10}{c}{With weights, $n=10{,}000$, effective $n=6{,}161$} \\
\multicolumn{10}{c} {public method: width $=$ 0.055, coverage $=$ 0.953, score $=$ 0.063} \\
0.2 &      0.055 &        0.523 &              0.618 &     0.173 &       0.938 &             0.226 &     0.173 &       0.940 &             0.226 \\
0.5 &      0.055 &        0.788 &              0.146 &     0.085 &       0.952 &             0.104 &     0.085 &       0.950 &             0.104 \\
1.0 &      0.055 &        0.909 &              0.076 &     0.064 &       0.959 &             0.074 &     0.064 &       0.957 &             0.074 \\
4.0 &      0.055 &        0.952 &              0.063 &     0.056 &       0.958 &             0.063 &     0.056 &       0.956 &             0.063 \\

\bottomrule
\end{tabular*}
\end{center}
\end{table}

\begin{table}[htbp]
\caption{Average width, coverage, and CI score of 95\protect \%
confidence intervals for the log ratio, using the Laplace mechanism. Public:
no noise added and thus Non-DP; the results do not change as a function
of $\epsilon $. No correction: ignoring the fact that DP noise was added.
Monte Carlo: correction via Monte Carlo simulation. Analytical: correction
via modified variance terms.}
\label{log_ratio_tab_laplace}
\renewcommand{\arraystretch}{1.05}
\begin{center}
\tabcolsep=0pt
\begin{tabular*}{\textwidth}{@{\extracolsep{\fill}}l ccc ccc ccc}
\toprule
    & \multicolumn{3}{c}{No Correction}
    & \multicolumn{3}{c}{Monte Carlo}
    & \multicolumn{3}{c}{Analytical} \\

   $\epsilon $  & width & coverage & Score
               & width & coverage & score
               & width & coverage & score \\
\midrule
\multicolumn{10}{c}{No weights, $n=5{,}000$, effective $n=5{,}000$} \\
\multicolumn{10}{c} {public method: width $=$ 0.055, coverage $=$ 0.953, score $=$ 0.066} \\
0.2 &      0.055 &        0.730 &              0.217 &     0.099 &       0.938 &             0.135 &     0.099 &       0.943 &             0.134 \\
0.5 &      0.055 &        0.902 &              0.086 &     0.064 &       0.946 &             0.081 &     0.065 &       0.946 &             0.081 \\
1.0 &      0.055 &        0.936 &              0.071 &     0.058 &       0.950 &             0.070 &     0.058 &       0.949 &             0.070 \\
4.0 &      0.055 &        0.948 &              0.067 &     0.056 &       0.951 &             0.067 &     0.056 &       0.950 &             0.067 \\[2ex]

\multicolumn{10}{c}{With weights, $n=5{,}000$, effective $n=2{,}622$} \\
\multicolumn{10}{c} {public method: width $=$ 0.071, coverage $=$ 0.948, score $=$ 0.085} \\
0.2 &      0.070 &        0.412 &              1.365 &     0.308 &       0.933 &             0.432 &     0.307 &       0.937 &             0.426 \\
0.5 &      0.071 &        0.696 &              0.329 &     0.138 &       0.941 &             0.184 &     0.138 &       0.944 &             0.183 \\
1.0 &      0.071 &        0.854 &              0.136 &     0.092 &       0.938 &             0.116 &     0.092 &       0.938 &             0.116 \\
4.0 &      0.071 &        0.940 &              0.087 &     0.072 &       0.947 &             0.087 &     0.072 &       0.947 &             0.087 \\[2ex]

\multicolumn{10}{c}{No weights, $n=10{,}000$, effective $n=10{,}000$} \\
\multicolumn{10}{c} {public method: width $=$ 0.039, coverage $=$ 0.948, score $=$ 0.046} \\
0.2 &      0.039 &        0.830 &              0.089 &     0.057 &       0.947 &             0.070 &     0.057 &       0.946 &             0.071 \\
0.5 &      0.039 &        0.933 &              0.051 &     0.043 &       0.953 &             0.050 &     0.043 &       0.953 &             0.050 \\
1.0 &      0.039 &        0.944 &              0.047 &     0.040 &       0.954 &             0.047 &     0.040 &       0.953 &             0.047 \\
4.0 &      0.039 &        0.949 &              0.046 &     0.039 &       0.949 &             0.046 &     0.039 &       0.949 &             0.046 \\[2ex]

\multicolumn{10}{c}{With weights, $n=10{,}000$, effective $n=6{,}161$} \\
\multicolumn{10}{c} {public method: width $=$ 0.050, coverage $=$ 0.951, score $=$ 0.057} \\
0.2 &      0.050 &        0.525 &              0.561 &     0.157 &       0.936 &             0.207 &     0.157 &       0.937 &             0.207 \\
0.5 &      0.050 &        0.789 &              0.133 &     0.078 &       0.950 &             0.095 &     0.078 &       0.950 &             0.095 \\
1.0 &      0.050 &        0.910 &              0.070 &     0.058 &       0.958 &             0.067 &     0.058 &       0.957 &             0.067 \\
4.0 &      0.050 &        0.953 &              0.057 &     0.050 &       0.956 &             0.058 &     0.050 &       0.956 &             0.058 \\

\bottomrule
\end{tabular*}
\end{center}
\end{table}

Under the same simulation settings other than switching to the Laplace
mechanism, Tables~\ref{ratio_tab_laplace} and~\ref{log_ratio_tab_laplace} show the same patterns as in Tables~\ref{ratio_tab} and~\ref{log_ratio_tab}: e.g., the no-correction method
under covers, and both the proposed methods have much better coverage.
In particular, although using the Laplace mechanism violates the assumption
of the Analytical method, where the numerator and denominator are no longer
Gaussians, the method's coverages are still close to the nominal level.
Also, compared with the Gaussian mechanism, smaller amounts of noise are
needed for the Laplace mechanism for the particular simulation setting,
which yields narrower CIs. If the number of computations increases though,\vadjust{\goodbreak}
for example in an application with multiple testing, the Gaussian mechanism
will start to provide narrower CIs. Practitioners are encouraged to compute
the variance of the noise under both mechanisms and choose the winner.

\end{appendix}

\begin{acknowledgement}
The authors are grateful to Imanol Arrieta Ibarra, Ilya Mironov, Jonathan
Tannen, Brian Karrer, and James Honaker for many discussions and for reading
earlier versions of the manuscript. The authors are grateful to the editor
and the reviewers, whose comments and suggestions have significantly improved
the paper.
\end{acknowledgement}

\bibliographystyle{jds}
\bibliography{References}

\end{document}